\documentclass[aip,pop,reprint,groupedaddress]{revtex4-1}

\usepackage{eurosym}
\usepackage{amssymb,amsmath}
\usepackage{graphicx}

\graphicspath{ {./figures/} }
\begin{document}

\title{Electron heating in 2-D:  combining Fermi-Ulam acceleration and magnetic-moment non-adiabaticity in a mirror-configuration plasma}
\author{C.P.S. Swanson}
\email{cswanson@pppl.gov}
\affiliation{Princeton Plasma Physics Laboratory, Princeton University, Princeton, New Jersey 08543, USA}
\author{C. A. Galea}
\affiliation{Princeton Fusion Systems, Plainsboro, New Jersey 08536, USA}
\author{S. A. Cohen}
\affiliation{Princeton Plasma Physics Laboratory, Princeton University, Princeton, New Jersey 08543, USA}
\date{\today}

\begin{abstract}
We analyze a new mechanism for the creation and confinement of energetic electrons in a mirror-configuration plasma. A Fermi-Ulam-type process, driven by  end-localized coherent electrostatic oscillations, provides axial acceleration while a natural non-adiabaticity of $\mu$ provides phase decorrelation and energy isotropization. This novel 2-D combination causes the electron energy distribution function, calculated with a diffusive-loss model, to assume a Maxwellian shape with the $\mu$  non-adiabaticity reducing loss-cone escape and annulling the absolute-barrier energy-limiting Chirikov criterion of lower dimensional models. The theoretical predictions are compared with data from an experiment. 
\end{abstract}

\maketitle

\address{\textit{Princeton Plasma Physics Laboratory, Princeton University,
Princeton, New Jersey 08543, USA}}

\section{Introduction}
\label{sec:bintro}

Charged particle confinement in axisymmetric mirror machines is  often justified by the assumption of magnetic moment, $\mu$, conservation. This leads to the concept of the mirror loss cone in which the ratio of the energy parallel to the magnetic field, \textbf{B}, to that perpendicular plays the decisive role. As this ratio drops, particle loss disappears at a critical value. It is for this reason that plasma heating in the  \textbf{B}-perpendicular direction, as by electron cyclotron resonance, is chosen for mirror machines.

In this paper we analyze the opposite situation -- with acceleration parallel to  \textbf{B} -- and show that particles can be heated to high energies and well confined even if particle collisions,  turbulence, or other common velocity-isotropization processes are not present. An essential contributor to the heating process  we  describe is the lack of $\mu$ conservation as particles traverse the mirror midplane.\cite{henrich_departure_1956,garren_individual_1958,tagare_motion_1986,zelenyi_quasiadiabatic_2013,hastie_non-adiabatic_1969,cohen_nonadiabaticity_1978-1}  The combination of energy and axial oscillation phase, and $\mu$ and gyrophase creates a two-dimensional coupled map. 

By itself, Fermi-Ulam-like acceleration $\it{via}$ weak electrostatic oscillations  cannot produce a Maxwellian-type electron energy distribution function, EEDF, particularly one that extends to high energies. \textbf{B}-parallel (axial) electrostatic oscillations  increase the parallel energy only, causing  particles to migrate into the loss cone. If the initial perpendicular energy were very high, the combination of oscillation amplitude and frequency and mirror-bounce transit time would limit energy gain $\it{via}$ the Chirikov criterion.\cite{Zaslavsky_fermi_1964, chirikov_stability_1978} We describe how both these apparent limitations are overcome by the natural non-adiabaticity of $\mu$ in mirror machines ascribed to a centrifugal kick near the mirror's axial midplane by the particle's axial velocity and mirror's radial field.\cite {delcourt_simple_1994}  

Section \ref{sec:fermi-ulam} contains the relevant characteristics of the Fermi-Ulam map. Section \ref{sec:beam} describes ways to create weak electrostatic oscillations near the mirror throats. This localization contributes to the similarity with Fermi-Ulam acceleration. Section \ref{sec:model} describes a particle's energy gain when it passes through an electrostatic oscillation. Section \ref{sec:K} describes a particle's long-time history from several such transits. We show that periodic forcing alone would not allow particles to traverse the Fermi-Ulam phase-space separatrix. Section \ref{sec:KPrime} describes magnetic moment ($\mu$) quasiadiabaticity in a magnetic mirror. Section \ref{sec:R} shows that $\mu$ quasiadiabaticity is sufficient to allow particles to circumvent the phase-space barriers of the Fermi-Ulam map. This section also discusses the lower limits for particle energy for which the necessary $\mu$-non-conservation will occur, which set the lower threshold of initial electron energy for further electron heating. Section \ref{sec:EEDF} describes the EEDF that results from Fermi-Ulam acceleration and decorrelation born of $\mu$ quasiadiabaticity. Section \ref{sec:experiment} compares experimental results with this model. 

\section{The Fermi-Ulam Map}
\label{sec:fermi-ulam}

The original second-order Fermi acceleration\cite{fermi_origin_1949} mechanism produces a power-law EEDF, $f(E)\propto E^{-r}$.  The Fermi-Ulam map considers a one-dimensional version of this process in which a particle bounces between two rigid fixed walls, one with an artificial sinusoidally varying velocity.\cite{ulam_statistical_1961} Instead of producing a power-law distribution, the Fermi-Ulam map shows  numerous adiabatic and one absolute barrier in phase-space, the latter leading to a finite-energy truncation of the EEDF.\cite{Zaslavsky_fermi_1964}

The Fermi-Ulam map can be reduced to the Standard (Chirikov-Taylor) Map:\cite{lieberman_stochastic_1972,lichtenberg_fermi_1980,karlis_hyperacceleration_2006}

\begin{equation}
p_{n+1}=p_n+K\sin(\theta_n) 
\label{eq:standard-map-p}
\end{equation}
\begin{equation}
\theta_{n+1}=\theta_n+p_{n+1}
\label{eq:standard-map-theta}
\end{equation}

\noindent where $K$ is the stochasticity parameter and $p,\theta$ are dimensionless degrees-of-freedom of the map. In the Fermi-Ulam case,  $p$ and $\theta$ correspond to the transit time and the oscillation phase of each bounce. 

 The existence of a stochastic sea in the Standard Map can be evaluated by the Chirikov Criterion, the change in oscillation phase upon return to the oscillating wall due to the   velocity increment imparted by the previous impact on the moving wall. If this value is larger than 1 radian of the wall's oscillation period, a stochastic sea exists and a particle is free to gain energy. If this value is less than 1 radian, particle orbits in phase space are quasi-periodic and the particle's energy is limited to a narrow region around its initial energy.\cite{chirikov_stability_1978} 

In the Fermi-Ulam map, $K$ is a decreasing function of velocity because, as the velocity increases, the transit time of a particle decreases, hence the oscillating wall has less time to change phase given a velocity increment. This means that a stochastic sea exists at low velocity but at a critical higher velocity, a separatrix exists and a particle cannot gain velocity above this value. This critical velocity is a function of the length between the walls and the strength and frequency of the forcing. 

Multiple alterations to the Fermi-Ulam map are known to destroy this separatrix. Some are: changing the sinusoidal forcing model to a sawtooth;\cite{lichtenberg_fermi_1980} changing the return-time function of velocity to one that is increasing rather than decreasing, $\it {e.g.}$, if the particle returned under gravity;\cite{lichtenberg_fermi_1980} and adding a random perturbation to the oscillation phase at each bounce.\cite{lieberman_stochastic_1972} In general, the addition of  dimensions to the dynamics destroys separatrices.\cite{chirikov_universal_1979}

\section{Localized $\textbf{B}_{||}$ electrostatic oscillations: applied $\it{vs}$ spontaneous}
\label{sec:beam}

For Fermi acceleration to energize particles parallel to \textbf{B}  in a mirror machine, a method to impart velocity increments must be employed. One method is to make  localized coherent electrostatic oscillations by placing near the mirror-machine throats a pair of closely spaced, parallel,  transparent metal grids with their surface normals parallel to \textbf{B}. These can be driven with voltage waveform shapes of controllable amplitudes and frequencies.

A spontaneous method invokes the  2-stream-instability mechanism suggested in a previous paper \cite{Swanson_2019} to explain experimental results in the PFRC-2 device.  In that experiment, measurements in one end cell of the mirror, the Far End Cell (FEC), showed a strongly negative  plasma potential, typically $-600$ V, while that in the mirror's central mirror cell (CC)  was near ground. This voltage drop accelerated a beam of nearly mono-energetic electrons from the FEC plasma into the CC. As described in Section VIII,  this beam-plasma system is expected to be unstable to longitudinal electrostatic modes. Such modes have been observed in double-layer experiments and attributed to a spontaneous beam-plasma 2-stream mechanism.\cite{schrittwieser_observation_1992} Probes in the PFRC-2 device detected electrostatic oscillations near a mirror throat in a frequency range (100-200 MHz) and of amplitude (50-150 V/cm), consistent with the 2-stream model.\cite{swanson_measurement_2018}

\section{Energy gain from a localized electrostatic fluctuation}
\label{sec:model}

This section describes two physical situations: An ``oscillating wall" case and a ``fixed wall with oscillation" case. The ``oscillating wall" case is that a particle gains or loses energy by bouncing off of a moving potential barrier. The oscillating wall case is commonly considered in the literature. 

The ``fixed wall with oscillation" case is that a particle gains or loses energy by bouncing off of a fixed potential barrier, with a smaller oscillating potential superimposed. The fixed wall with oscillation case has some important differences with the oscillating wall case, and more accurately represents the process occurring in PFRC-2. The fixed wall is the static magnetic mirror potential, and the oscillating potential is an electrostatic oscillation that occurs near the mirror nozzle. 

Consider the following two soft-wall effective potentials, Equations \ref{eq:fermi-wall-potential} and \ref{eq:fermi-magnet-wall-potential}. 

Equation \ref{eq:fermi-wall-potential} represents the oscillating wall case, in which a potential barrier ($U_{wall}$) is oscillating axially. Equation \ref{eq:fermi-wall-potential} can be shown to reduce to the Fermi-Ulam map in certain limits.

\begin{equation}
U_{wall}(x,t)=E_0 \text{e}^{-(x- \int v_w dt)/{x_c}}
\label{eq:fermi-wall-potential}
\end{equation}

\noindent where $v_w(t)=v_{w,0}\sin(\omega t)$ is the oscillation ``velocity" of the wall,  $v_{w,0}$ is the pre-factor of the velocity, $\omega$ is the wall's oscillation frequency, and  $x_c$ is the characteristic distance of the soft-wall  potential fall. $x = 0$ is the reflection location at the particle's initial energy, $E_0$.

Equation \ref{eq:fermi-magnet-wall-potential} is the fixed wall with oscillation case, in which the particle bounces back from a stationary potential barrier with a small oscillating potential superimposed.

\begin{equation}
U_{pert}(x,t)=E_0 \text{e}^{-x/x_c}+E_1 \text{e}^{(x-\int v_w dt)/x_{c}}
\label{eq:fermi-magnet-wall-potential}
\end{equation}

The first RHS term of $U_{pert}$ corresponds to a static confining potential, such as created by a mirror's throat. The second term is a small added moving electrostatic perturbation of strength  $E_1$.  In this analysis we assume $E_1/E_0 \ll 1$.   

A particle incident on these potentials from the right ($+x$) with some velocity $v_p$ will bounce back. It may gain or lose energy dependent on the oscillation phase. This energy can be computed in the limit $1/\omega \gg t_{r}$, where $t_{r} = x_c/v_{p}$ is the approximate interaction time of the particle with the ramp. The maximum energy gain with which a particle bounces back from an oscillating wall (Equation \ref{eq:fermi-wall-potential}) is:

\begin{equation}
\Delta E_{wall}=4 \sqrt{\frac{1}{2}m_e v_w^2} \sqrt{E_0} 
\label{eq:Delta-fermi-ulam}
\end{equation}

In the further limit that $t_{t} \gg 1/\omega$, where $t_{t} = L/v_{p}$, the approximate transit time between the ramps, Equation \ref{eq:Delta-fermi-ulam} reduces to the Fermi-Ulam map.

In contrast the fixed wall with oscillation (Equation \ref{eq:fermi-magnet-wall-potential}) in the same limt yields a maximum energy gain of

\begin{equation}
\Delta E_{perturb}=4\sqrt{\frac{1}{2}m_e v_w^2}\frac{E_1}{\sqrt{E_0}},
\label{eq:Delta-E}
\end{equation}

\noindent a result clearly different than Equation \ref{eq:Delta-fermi-ulam}. For the oscillating wall case (Equation \ref{eq:Delta-fermi-ulam}), the particle gains more energy from the bounce when it is incident with more energy. In contrast, for the fixed wall with oscillation case (Equation \ref{eq:Delta-E}), the particle gains \textit{less} energy from the bounce when it is incident with more energy. This can be thought of in the following way: higher-energy particles spend less time in the area of interaction than lower-energy particles, and hence their energy changes less. 

Equations \ref{eq:fermi-wall-potential} and \ref{eq:Delta-fermi-ulam} are included for comparison to the Fermi-Ulam map. For the PFRC-2 experiments described, the case represented by Equations \ref{eq:fermi-magnet-wall-potential} and \ref{eq:Delta-E} is expected to be closer to the physical situation.

\section{The energy trajectory resultant from many such bounces}
\label{sec:K}

In the last section, Section \ref{sec:model}, we analyzed the effect of a single bounce on the energy of a particle. In this section, we analyze the effect of many successive bounces on the energy of a particle. 

Essential to this analysis is the transit time of the particle, $t_t$. After a particle bounces off of the mirror nozzle, a time $t_t$ elapses before the particle is once again incident on that nozzle. $t_t$ can be computed:

\begin{equation}
t_t(E,\mu) = 4 \int_0^{z_l} \frac{dz}{\sqrt{2E/m - 2\mu B(z) /m}},
\label{eq:transit-time-mu}
\end{equation}

\noindent where $z_l$ is the turning point of the particle at the mirror nozzle, $z$ is the axial distance, $E$ is the particle's energy, $m$ is the particle's mass, $\mu$ is the particle's magnetic moment, and $B(z)$ is the strength of the magnetic field at axial point $z$.

The long-time trajectory of the particle in energy can be shown to reduce to the standard map, repeated here from Equations \ref{eq:standard-map-p} and \ref{eq:standard-map-theta}:

\begin{equation*}
p_{n+1}=p_n+K\sin(\theta_n)
\end{equation*}
\begin{equation*}
\theta_{n+1}=\theta_n+p_{n+1}
\end{equation*}

$p_n$ is the product of the transit time of the particle and the potential oscillation frequency $\omega t_t$ on the $n$th bounce. $\theta_{n}$ is the oscillation phase of the potential on the $n$th bounce. $K$ is a parameter related to the magnitude of the energy increments that occur during a bounce:

\begin{equation}
K=\partial_E p \Delta E,
\label{eq:K1}
\end{equation}

In words, $K$ is the amount that the increment in energy is able to change the potential oscillation phase when the particle is next incident on the oscillating potential region. 

We may use the magnitude of $K$ to determine whether it is possible for a particle to gain energy up to and beyond 30 keV, as is measured in the PFRC-2.\cite{Swanson_2019} The criterion $K>1$ is called the Chirikov Criterion.\cite{chirikov_stability_1978}

If $K>1$, chaos exists in the 1-D map and particle energies are free to diffuse (gain energy without limit). If $K<1$, phase space separatrices exist in the 1-D map and particles' energies are kept in quasiperiodic orbits in the vicinity of the original particle energy. If a region for which $K<1$ abuts a region for which $K>1$, a particle from the $K < 1$ region may diffuse in energy up to the critical energy which separates the regions, but no farther. 

\begin{figure}[tbp]
\centering\includegraphics[scale=0.32]{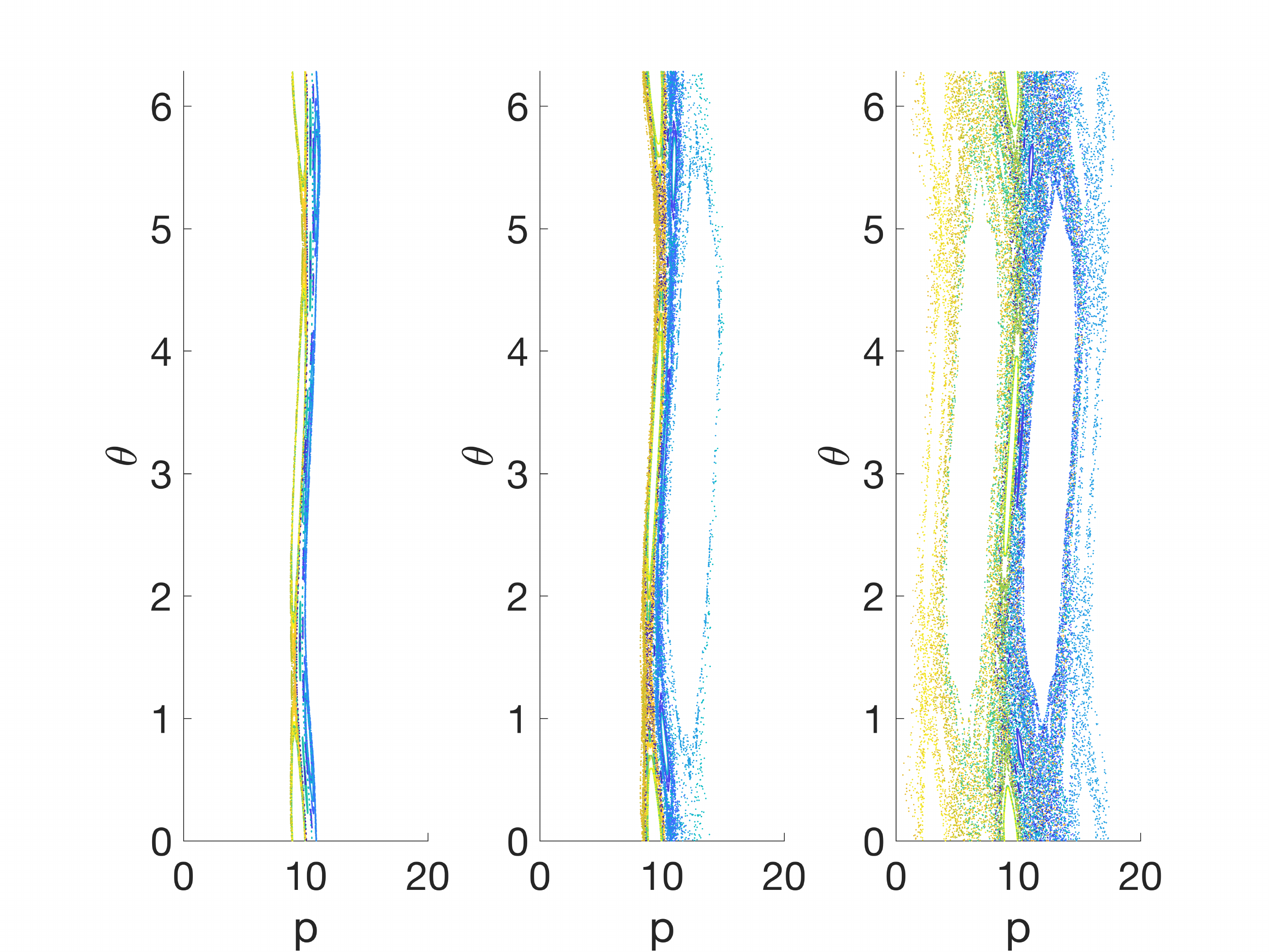}
\caption{Plots of $p,\theta$ points produced by applying Equations (\ref{eq:standard-map-p}) and (\ref{eq:standard-map-theta}) to 20 points originating at evenly spaced $\theta$ values and $p = 10$. $K$ = 0.9, 1.1, and 1.3. Color describes the initial $\theta$ of the point. 2000 time steps were performed.}
\label{fig:CC-chaos}
\end{figure}

Figure \ref{fig:CC-chaos} shows three cases to illustrate this point: $K=0.9$ in which particles are contained to a narrow region; $K=1.1$ in which particles diffuse freely; and $K=1.3$ in which particles diffuse very quickly. In Section VIII we shall show that $K \sim 0.1$ for electrons of the relevant energy of 3 keV in the PFRC-2.  Therefore there must be some other phenomenon which allows PFRC-2 particles to cross the purported separatrices between different energies. In Section \ref{sec:R}, we will show that, whether or not it is the only such phenomenon, the non-adiabaticity of $\mu$ is a sufficient phenomenon to allow this. 

\section{The quasiadiabatic behavior of electrons in the PFRC-2}
\label{sec:KPrime}

It does not take extreme field curvature or a magnetic null to produce large changes in $\mu$, a fact  known since the 1950s\cite{henrich_departure_1956,garren_individual_1958} and significantly explored since. Publications describe both the action of a single pass into a non-adiabatic region\cite{hastie_non-adiabatic_1969,dykhne_change_1960,cohen_nonadiabaticity_1978} and the compounded effect of many such passes.\cite{chirikov_stability_1978,tagare_motion_1986,zelenyi_quasiadiabatic_2013} The name ``quasiadiabaticity" is given to the case that particles' $\mu$ are well-conserved for the majority of their trajectories, but pass through specific regions where their $\mu$ undergo a discrete change in value.\cite{zelenyi_quasiadiabatic_2013}

\begin{figure*}[tbp]
\centering\includegraphics[scale=1]{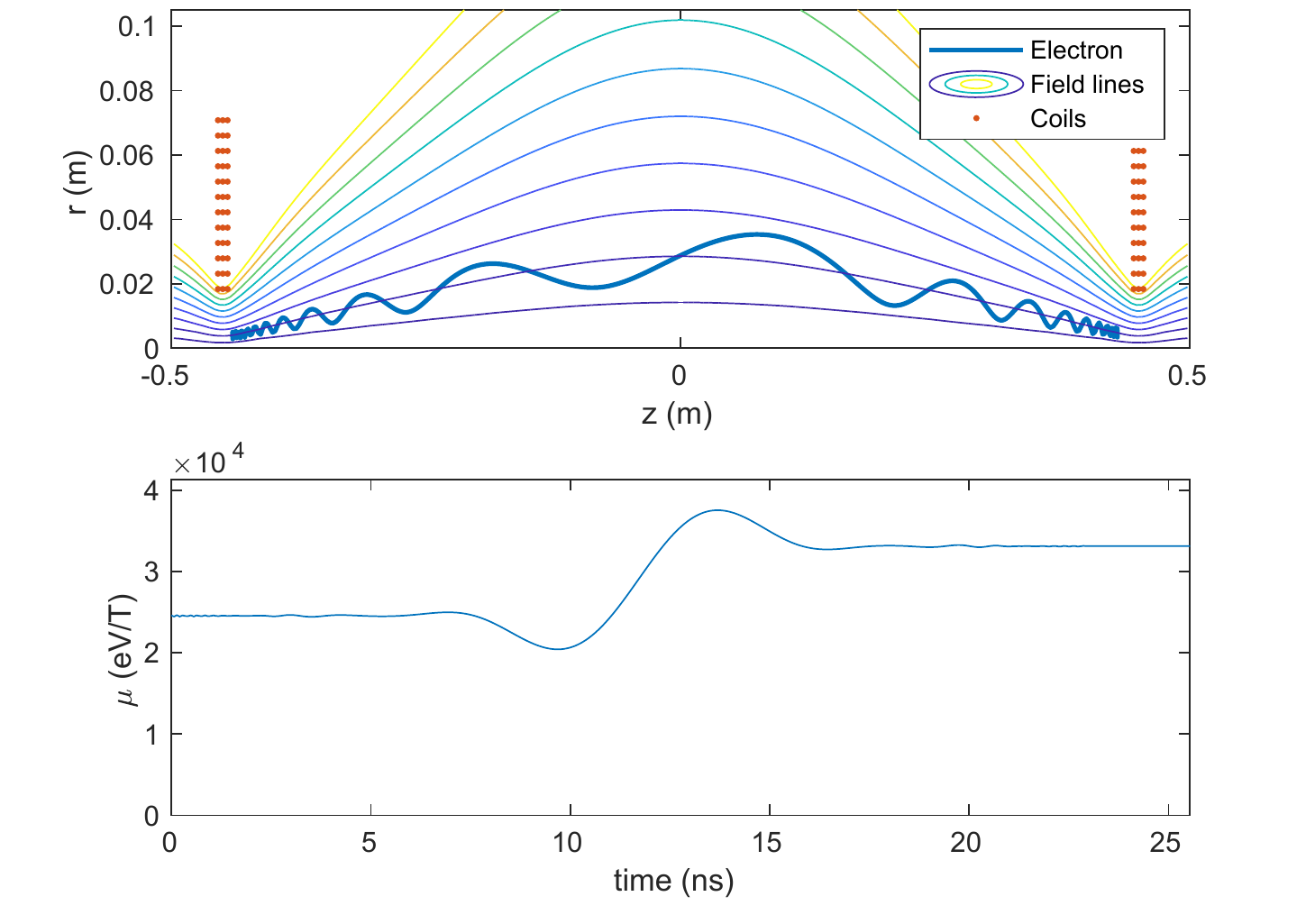}
\caption{Boris-algorithm numerically calculated trajectory of a single electron in the PFRC-2. Top: the electron's  trajectory in space superimposed over the mirror geometry. Bottom:  $\mu$ \textit{vs.} time. The 5.4 keV electron starts in the mirror throat, marginally trapped and at 4-mm radius. In its ballistic trajectory, this electron gains 30\% of its initial $\mu$ on crossing the $z=0$ midplane.}
\label{fig:CC-Boris-Single}
\end{figure*}

A $keV$ electron in the PFRC-2, simply following its ballistic trajectory collisionlessly, \textit {i.e.}, without particle-particle collisions, and starting marginally trapped at a medium radius, 6 mm at the nozzle, may readily gain or lose 50 \% of its $\mu$.  Figure \ref{fig:CC-Boris-Single} shows such behavior for a 5.4 keV electron. Calculated $\mu$ changes well reproduce the approximate formulae of Hastie, Taylor, and Hobbs.\cite{hastie_non-adiabatic_1969} It is worth noting that the  \textit{traditional} adiabatic parameter, $\epsilon = \rho \nabla B/B$, is small, \textit{ca.} 0.01, and that the \textit{true} adiabatic parameter includes contributions from the parallel velocity and the second derivative of the curvature of the magnetic field lines.

This change in $\mu$ is dependent on the gyrophase at the midplane. Similarly to the oscillation phase 1-D map mentioned in Section \ref{sec:K}, the $\mu$ of a particle also follows a 1-D map which reduces to the standard map, but with different definitions for $(p,\theta)$ in Equations \ref{eq:standard-map-p} and \ref{eq:standard-map-theta}. The theory of multiple $\mu$ non-adiabatic changes is what gives the standard map its original name, the Chirikov Map.\cite{chirikov_stability_1978} 

In Section \ref{sec:K}, to determine the long-time behavior of the energy $E$, we started by determining the difference in the potential oscillation phase between successive increments to the energy, $\omega t_t$ (Equation \ref{eq:transit-time-mu}). In this section, to determine the long-time behavior of $\mu$, we start by determining the difference in the midplane gyrophase between successive increments to the $\mu$:

\begin{equation}
w_{n+1}=w_n+K'\sin(z_n) 
\label{eq:standard-map-w}
\end{equation}
\begin{equation}
z_{n+1}=z_n+w_{n+1}
\label{eq:standard-map-z}
\end{equation}

$z_n$ is the gyrophase when the particle crosses the midplane the $n$th time. $w_n$ is the difference between gyrophases at successive midplane crossings:

\begin{equation}
w = 2 \frac{e}{m} \int_0^{z_l} \frac{B(z)dz}{\sqrt{2E/m - 2\mu B(z)/m}}
\label{eq:psi0}
\end{equation}

\begin{equation}
K'=\partial_\mu w \Delta\mu
\end{equation}

\noindent where $\Delta\mu$ is the characteristic increment to $\mu$ in one transit of the machine. 

\subsection{A note on quasiadiabatic electrons in magnetic mirror machines}

The system defined by Equations \ref{eq:standard-map-w}, \ref{eq:standard-map-z}, and \ref{eq:psi0} are worth studying in their own right. Traditionally, the velocity space of particles in a magnetic mirror is split into two regions: The loss cone of passing particles, for which $\mu<\mu_p$, and the region of trapped particles, for which $\mu>\mu_p$. A close examination of these equations reveals that there are actually three regions of velocity space. Another critical value, $\mu_c$, defined as that $\mu$ for which $K'=1$, divides the adiabatically trapped region into two regions. $\mu>\mu_c$ is trapped as before; however, $\mu_p<\mu<\mu_c$ is an interesting region in which particles' $\mu$ are free to diffuse. The particles' $\mu$ may diffuse as high as $\mu_c$, and as low as $\mu_p$. The particles' $\mu$ can not diffuse higher than $\mu_c$. If the particle's $\mu$ diffuses lower than $\mu_p$, the particle will be lost on the next bounce. 

\begin{figure}[tbp]
\centering\includegraphics[scale=0.62]{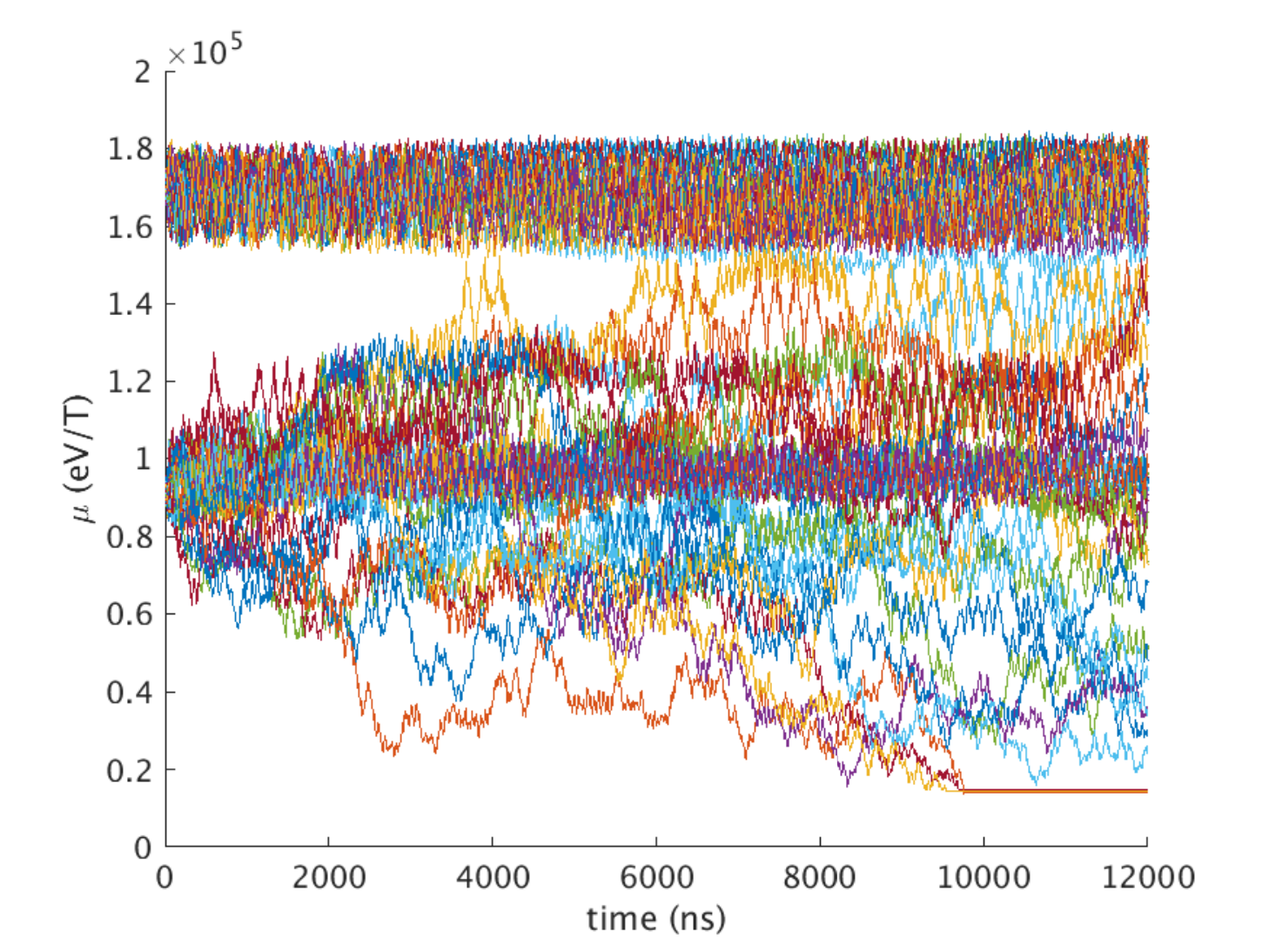}
\caption{Boris-algorithm simulations of $\mu$ trajectories for two particle ensembles in the PFRC-2. One 32-particle ensemble   was initialized with $\mu = 10\mu_p$ and the other with $6\mu_p$. A boundary is apparent between $\mu$ trajectories of the two ensembles. Particles with constant $\mu$ beyond 10000 ns have been lost.} 
\label{fig:CC-Boris-Separatrix}
\end{figure}

Because particles with $\mu<\mu_c$ may collisionlessly leave the mirror, this region can be thought of as an extended loss cone, which takes several bounces to leave. To test this behavior, we have conducted two Boris-algorithm ensemble simulations, one at a larger $\mu$ than the observed $\mu_c$, and one at a smaller $\mu$ than the observed $\mu_c$, see Figure \ref{fig:CC-Boris-Separatrix}. These are 3.6 keV particles, at a radius for which their $\mu_c$ is predicted to be equal to $10.7\mu_p$. As depicted in Figure \ref{fig:CC-Boris-Separatrix}, the smaller $\mu$ ensemble does exhibit an upper boundary $\mu$ beyond which it cannot diffuse. The observed $\mu_c$ is close to $8.8\mu_p$. This discrepancy may be due to the inexact nature of the Hastie, Taylor, and Hobbs formulae. 

Interestingly, it appears that there are \textit{always} particles with $\mu_p<\mu<\mu_c$, no matter how strong and smooth the magnetic field. Observe that, at $\mu_p$, the integral in Equation \ref{eq:psi0} diverges and $K'\rightarrow \infty$. Thus there is always some $\mu_c>\mu_p$ for which $K'=1$, even though this region may be extremely narrow. 

For the case of the PFRC-2, it is likely that nearly 100\% of the particles accelerated by the electrostatic potential have $\mu_p<\mu<\mu_c$. This is because these particles begin in the SEC, and so when they enter the CC they are by definition passing. Only those particles whose $\mu$ diffuse into $\mu_p<\mu<\mu_c$ persist for an appreciable time (many bounces). 

\section{The coupled map of $E,\theta,\mu,z$}
\label{sec:R}

In Section \ref{sec:K}, we established that an electrostatic oscillation near one nozzle with the amplitude measured could not heat particles to the 30+ keV measured in the PFRC-2. However, Equation \ref{eq:K1} assumes perfect adiabaticity of $\mu$. In Section \ref{sec:KPrime}, we found that the $\mu$ of these particles is extremely mobile. This chaotic $\mu$ behavior may be used to explain how the Fermi-accelerating electrons may circumvent their $K<1$ limit and become heated to very high energies.

By using the formulae of Hastie, Taylor, and Hobbs and Equation \ref{eq:transit-time-mu}, we may evaluate an equivalent Chirikov parameter for the effect of $\mu$ on potential oscillation phase:

\begin{equation}
R=\partial_\mu p \Delta\mu
\end{equation}

\noindent where $p$ is the difference in potential oscillation phase between successive mirror bounces, $\omega t_t$, where $t_t$ is defined in Equation \ref{eq:transit-time-mu}. $\omega \approx 2\pi\times 200$MHz is the angular frequency of the oscillation (the choice of 200 MHz is described in Section \ref{sec:experiment}), and $\Delta \mu$ is evaluated using the formulae of Hastie, Taylor, and Hobbs. For the PFRC-2 magnetic field and 3 keV electrons starting at a medium radius, $R$ is numerically evaluated to be $\sim 0.1$, of the same order as $K$.

As the increment in $\mu$ is dependent on gyrophase $w$ and the increment in $E$ is dependent on oscillation phase $\theta$, the increment in $\theta$ due to $\mu$ can be modeled as a coupling between two Chirikov maps, ($p$, $\theta$) and ($w$, $z$):

\begin{equation}
p_{n+1}=p_n+K\sin(\theta_n)+R\sin(z_n)
\label{eq:coupled-map-p}
\end{equation}
\begin{equation}
\theta_{n+1}=\theta_n+p_{n+1}
\label{eq:coupled-map-theta}
\end{equation}
\begin{equation}
w_{n+1}=w_n+K'\sin(z_n) 
\label{eq:coupled-map-w}
\end{equation}
\begin{equation}
z_{n+1}=z_n+w_{n+1}
\label{eq:coupled-map-z}
\end{equation}

\noindent where, recalling from Sections \ref{sec:K} and \ref{sec:KPrime}, $\theta_n$ is the oscillation phase of the electrostatic oscillation when the particle is incident on the oscillating region for the $n$th time, $p_n$ is the transit time of the particle multiplied by omega on the $n$th bounce (the number of oscillation periods that elapse), $z_n$ is the gyrophase of the particle at the midplane on its nth transit, and $w_n$ is the integral of the transit time multiplied by the local gyrofrequency (the number of gyroperiods that elapse) over the $n$th transit of the machine. 

Typical values of $K,K'$ for hot electrons in the PFRC-2 are: $K\sim 0.1$, per Section \ref{sec:K}, and $1<K'<\infty$, per Section \ref{sec:KPrime}. In later paragraphs, we use $K'=2.5$ for illustration purposes. In the case of uncoupled maps ($R=0$), these $K$ values would imply the ($p, \theta$) map is stable while the ($w, z$) map is unstable. The coupling appears as the last term in Equation \ref{eq:coupled-map-p} and was chosen to be unidirectional from ($w, z$) $\to$ ($p, \theta$) since the ($w, z$) map exhibits chaos without the need for coupling. It is this coupling term that is responsible for producing and destroying separatrices in the ($p, \theta$) map.

\begin{figure}[tbp]
\includegraphics[scale=0.32]{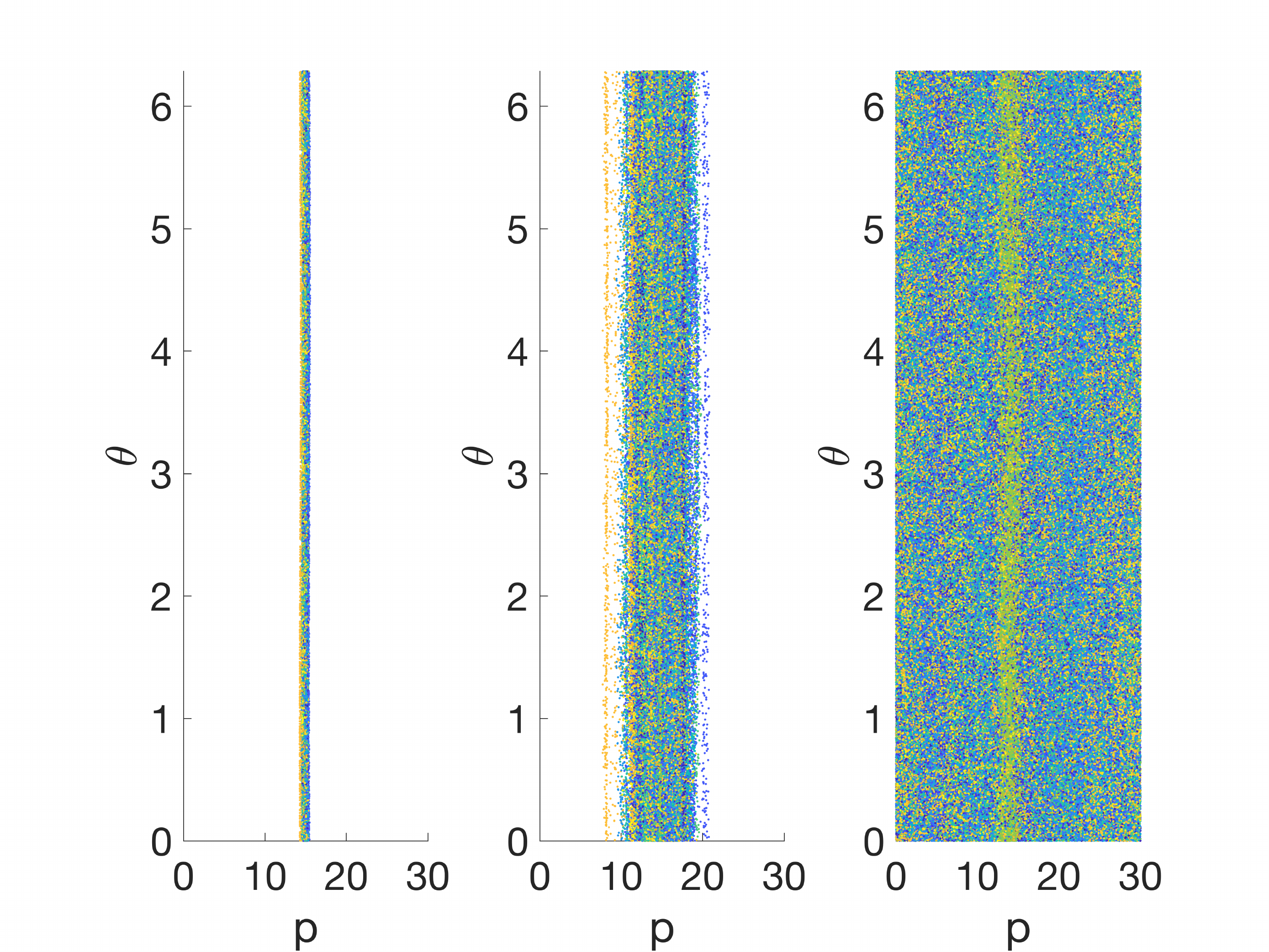}
\caption{Plots of $p, \theta$ from the coupled map defined by Equations \ref{eq:coupled-map-p}-\ref{eq:coupled-map-z}. Each plot is for $K=0.1$, $K'=2.5$. From left to right, these plots have $R=0.01, 0.1,$ and $1$. 1600 time steps are depicted. 400 points, evenly spaced in $\theta$, were initialized at $p=15$. The color of the points corresponds to the initial $\theta$ values.}
\label{fig:fermi-chaos}
\end{figure}

To verify this, we performed numerical iteration of the coupled map defined by Equations \ref{eq:coupled-map-p}-\ref{eq:coupled-map-z}. Results for the ($p,\theta$) map are plotted in Figure \ref{fig:fermi-chaos}. The case that $R = 0.1$ shows diffusion beyond the quasiperiodic initial orbits, and the case that $R = 1$ shows much faster diffusion. Thus, while the Chirikov criterion alone indicates that electron energy is constrained to vary only within a narrow band around the initial energy for small $K$, the natural, collisionless changes in $\mu$ in the PFRC-2 are sufficient to allow diffusion. 

$\mu$ changes must be greater than $\sim 10\%$ to markedly enhance energy diffusivity. The Hastie, Taylor, and Hobbs formula prediction for the PFRC-2 is that the minimum required energy at $B(0,0)= 60$ G is about 1 $keV$. For these parameters, the ratio of the electron gyroradius to the field curvature, the traditional adiabatic parameter,  at $ r = 8 $ cm of the PFRC-2 midplane is $\sim 0.003$. As observed in the PFRC-2, a low gas pressure, below $\sim 0.5$ mT, is required for gas excitation and  ionization to not act as large drains on the energy gain. In the PFRC-2, the source of $keV$ electrons is the capacitively coupled plasma in the Source End Cell (SEC). The EEDF in this SEC plasma satisfies\cite{jandovitz_demonstration_2018} the $keV$ requirement, hence the strong dependence of the high energy X-ray flux, the proxy for high energy electrons, on SEC RF power. 

\section{EEDF evolution equation}
\label{sec:EEDF}

In this section, we will assume that the phases of the electrostatic oscillation each time a particle is incident are decorrelated. As we discussed in Section \ref{sec:R}, the non-adiabaticity of $\mu$ is sufficient to provide this decorrelation. 

The action of random increments to the energy is diffusive, 

\begin{equation}
\partial_t f = \partial_E D_E \partial_E f
\end{equation}

\noindent where $f$ is the particle distribution function, $D_E$ is the energy diffusivity, $D_E=\langle \Delta E^2\rangle/t_{t}$, $t_{t}$ is the transit time between energy increments, and $\partial_i$ is the partial derivative with respect to the variable $i$. 

Other effects assumed to be important to shaping the EEDF are particle loss rate, $-f/\tau$, where $\tau$ is the particle loss time, and energy loss, $-(\partial_t E)\partial_E f$, where $\partial_t E$ is the energy loss rate of a fast electron.

\begin{equation}
\partial_t f = \partial_E D_E \partial_E f-(\partial_t E)\partial_E f-f/\tau
\end{equation}

Effects that might contribute to the middle term include (gas) ionization or excitation and X-ray emission. In steady state and far in energy from any sources of particles, the Green's function solution to this equation is

\begin{equation}
f\propto \text{e}^{-T_{eff}},
\end{equation}

\noindent where

\begin{equation}
T_{eff}=\sqrt{D_E \tau + \frac{1}{4}(\partial_t E \tau)^2} + \frac{1}{2}\partial_t E \tau.
\end{equation}

In the  limit that $\partial_t E \tau\ll \sqrt{D_E \tau}$, expected at low gas pressure, plugging in the definition of $D_E$ yields

\begin{equation}
T_{eff} = \Delta E \sqrt{\frac{\tau}{t_{t}}} + \frac{1}{2} \partial_t E \tau.
\label{eq:Teff}
\end{equation}

\section{Comparison with an experiment}
\label{sec:experiment}

Pulsed, high power ($10^4 - 10^{10}$ W) electron beams injected into magnetic mirrors have been used to create microsecond-duration high temperature ($>$10 keV), high density ($>10^{13}$ cm$^{-3}$)  plasmas,  relevant to beam-plasma interaction,\cite{smullin_generation_1962,alexeff_hot-electron_1963,alexeff_beam-plasma_1964} electrostatic turbulence,\cite{demirkanov_plasma_1965,blinov_plasma_1967,alexeff_oscillations_1968} atomic physics processes,\cite{alexeff_observation_1964} and nuclear fusion.\cite{alexeff_production_1967,arzhannikov_new_1988,burdakov_plasma_2007} These plasmas are generally observed to have Maxwellian electron energy distribution functions (EEDF) and turbulent electrostatic wave spectra. The accepted mechanism for electron heating is turbulent electrostatic heating along the beam column.

In contrast, in recent studies \cite{Swanson_2019,jandovitz_demonstration_2018} the PFRC-2 device was run as a steady-state magnetic mirror. Plasma was formed and heated by 50 - 500 W of capacitively-coupled RF power. Run in this mode, the PFRC-2 has more in common with a low-temperature plasma apparatus than with the high-power electron-beam heating experiments. Even so, a ``hot'' minority component having $n_e\approx 3\times10^7 \textrm{ cm}^{-3}$ and $T_e\approx3 \textrm{ keV}$ was observed  in the PFRC-2 central cell (CC). Based on X-ray measurements, some electrons had energies exceeding 30  keV. 

A previous paper reported on the measurement in the PFRC-2 SEC of a warm minority component having $n_e\approx 3\times10^8 \textrm{ cm}^{-3}$ and $T_e\approx300 \textrm{ eV}$.\cite{jandovitz_demonstration_2018} That paper also described a near-$kV$ potential difference between the far end cell, FEC (negative), and the CC (near ground), considerably higher than commonly found in double layer devices.\cite{charles_review_2007} That potential  spontaneously generates  a nearly monoenergetic beam of electrons that propagates from the FEC into the CC. 

The parameters of this beam are proper to generate a two-stream instability, creating  electrostatic oscillations that are localized in the CC near the FEC. The amplitude of these oscillations, measured to be near 50 Volts, is  two orders-of-magnitude lower  than the turbulent electrostatic waves in the aforementioned energetic-beam mirror discharges. Moreover, their  spectra  are sinusoids and harmonics thereof. Nonetheless, in Ref.[\cite{Swanson_2019}] we reported on measurements of a minority ``hot'' component with a near-Maxwellian EEDF and attributed this population to the low amplitude coherent oscillations. The low value of all potentials (the DC space potentials and the electrostatic oscillations) is very low compared to the 10's of keV electron energies. This, plus the Maxwellian-like shape of the EEDF, necessitated our consideration of a different mechanism of electron heating. In Ref.[\cite{Swanson_2019}], we presented a heuristic model based on a modified multi-dimensional Fermi-Ulam acceleration process. Herein, we use the methods described in Sections \ref{sec:beam}-\ref{sec:EEDF} to explain the experimental results.

\begin{figure}[tbp]
\centering\includegraphics[scale=0.58]{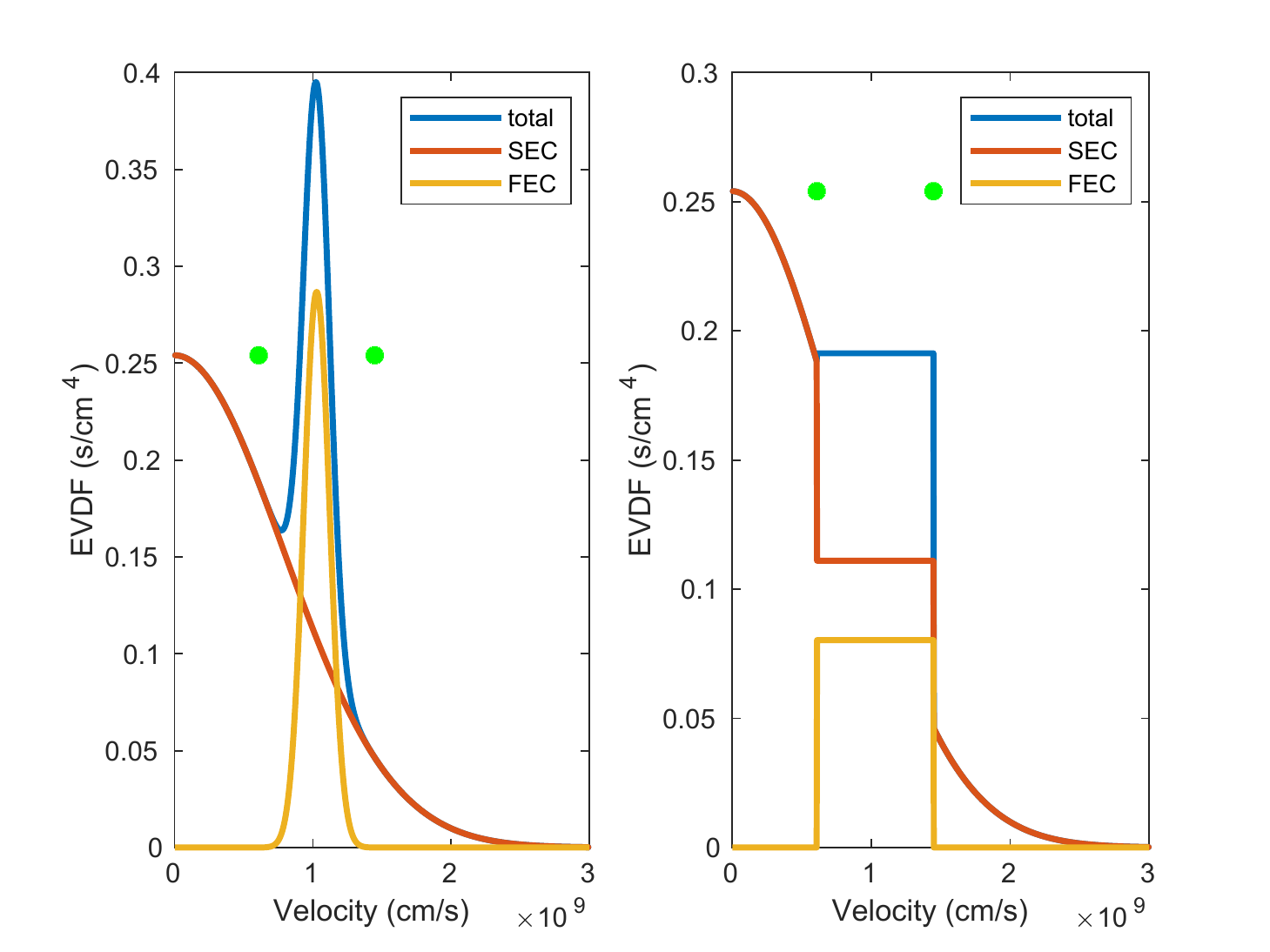}
\caption{Left: EVDFs of the warm and beam electrons given in the text, and their sum. This EVDF is unstable to electrostatic modes. Right: EVDF of the same plasma, except that the EVDF has been flattened in the velocity region that corresponds to a 50 $V_{pkpk}$ electrostatic oscillation. This oscillation is sufficient to make $f'(v)\le0$ everywhere, the condition for two-stream instability saturation in the inverse Landau damping limit. The bulk electron EVDF is not shown.}
\label{fig:fermi-landau-saturation}
\end{figure}

For specificity to the beam-plasma 2-stream instability question, we choose to evaluate the measurement-informed case \cite{Swanson_2019} that the beam electrons have $n_e=3.5\times10^7/\text{cm}^3,$ a 300 eV drift energy and an effective temperature of 5 eV. This electron velocity distribution function, EVDF, is depicted in Figure \ref{fig:fermi-landau-saturation}.

We numerically evaluated the  Nyquist theorem criterion for electrostatic mode stability and found this EVDF to be unstable.\cite{stix_waves_1992} In the inverse Landau damping limit, the instability condition $f'(v)>0$ is also clearly satisfied.

The expected saturation amplitude of this oscillation is calculable from a kinetic model.\cite{dewar_saturation_1973} This model roughly agrees with the inverse Landau damping limit of the saturation condition, that $f'(v)$ is nowhere positive.

By taking the EVDF of the warm electrons as linear around the velocity of the entering beam electrons, we may derive an approximate equation for the amplitude of the oscillation:

\begin{equation}
eV_{pkpk}\approx T_{warm} \frac{1}{2}\frac{n_{beam}}{n_{warm}} \sqrt{\frac{\pi T_{warm}}{E_{beam}}} \text{e}^{\frac{E_{beam}}{T_{warm}}}
\label{eq:fermi-Vpkpk}
\end{equation}

where $V_{pkpk}$ is the peak-to-peak saturation voltage of the oscillation, $T_{warm},n_{warm}$ are the temperature and density of the warm electrons, and $E_{beam},n_{beam}$ are the drift energy and density of the beam electrons. We expect our specific warm-beam plasma system to saturate at 50 $V_{pkpk}$ according to Equation \ref{eq:fermi-Vpkpk}. This is consistent with the measurement made in Ref.[\cite{Swanson_2019}].

The measured spectrum of electrostatic oscillations showed a broad peak around 200 MHz, close to the plasma frequency of the warm population, but it is also close to the cyclotron frequency of electrons at the measured point. Magnetic oscillations were shown to be absent at levels above 0.1 G, that is, the measured signal is purely electrostatic. Nevertheless, the precise nature of the instability is not yet understood.\cite{gary_electrostatic_1985} 

Expanding Equation \ref{eq:Teff} using the definition of $\Delta E$ $\it{via}$ Equation \ref{eq:Delta-E},

\begin{equation}
T_{eff} = 4 e\tilde{V} \sqrt{\frac{E_{beam}}{E}} \sqrt{\frac{\tau}{t_t}} + \frac{1}{2} \partial_t E \tau
\label{eq:effective-temperature}
\end{equation}

Evaluating $T_{eff}$ at an example energy of 3 keV, we find that the oscillation amplitude $\tilde{V}=25$ V, $E_{beam}=300$ eV, $E$=3 keV, $\tau(3\text{keV})$ was measured to be 150 $\mu$s in our companion paper, $t_t(3\text{keV})=25$ ns is calculable from the dimensions of the machine, and $\partial_t E = -8.4 \times 10^6$ eV/s is calculable from the NIST ionization cross sections.\cite{kim_NIST_2004} 
This produces a $T_{eff}=1.8$ keV, close to the measured value. 

Moreover, evaluating $K$ for the measured forcing of the PFRC-2 is possible. Recall from Section \ref{sec:R} that $\alpha=t_t \omega$, $\omega \approx 2\pi\times 200$MHz, and transit time $t_t$ is defined in Equation \ref{eq:transit-time-mu}, which can be substituted into Equation \ref{eq:K1} to obtain $K$. The measured forcing in the PFRC-2 yields $K\sim 0.1$, insufficient to allow energy diffusion, supporting our claim that non-conservation of $\mu$ is the cause of the needed de-correlation. 

In our companion paper, we verify the expected dependence on $\tilde{V}$ by increasing the neutral gas density in the FEC and so increasing beam current. The expected linear relationship between $\tilde{V}$ and $T_{eff}$ is measured. 

We also verified the expected dependence on $\tau$ by increasing the neutral gas density in the CC,  increasing collision-induced particle loss. Agreement was again obtained between the measured $T_{eff}$ and the measured $\tau$.

Because of the agreement between Equation \ref{eq:effective-temperature} and the measured temperatures in our companion paper, we propose the diffusion of particle energy under the influence of a spontaneous two-stream electrostatic instability as the mechanism for accelerating electrons to the high temperatures seen in the PFRC-2. 

\section{Summary}

In this paper, we have described a novel plasma heating process which we believe to be heating warm electrons to 3 keV temperatures in the PFRC-2, as measured in a companion publication.\cite{Swanson_2019} It is Fermi acceleration from a localized, sinusoidal, electrostatic fluctuation. This arises from two-stream instability from a spontaneously generated beam, caused by ionization downstream of a potential drop. We have given a simple model for the amplitude of the oscillation based on the inverse-Landau-limit saturation criterion, $f'(v)<0$. We have shown via a diffusive-loss model that this localized oscillation, combined with the natural motion of the particles in the magnetic mirror field, causes the particles to assume a roughly Maxwellian EEDF with a predicted temperature which agrees with the measured. 

However, periodic forcing alone would not allow acceleration to the high energies observed in the PFRC-2, due to the existence of phase-space separatrices in maps which reduce to the Standard Map. We have shown a sufficient phenomenon to break this separatrix, the phase-decorrelation effect of the natural non-adiabatic mobility of the magnetic moment, $\mu$. We showed this by implementing a coupled map and by a numerical iteration thereof. Finally, we have presented evidence that this same non-adiabaticity of $\mu$ could be leading to another anomalous measurement in the PFRC-2, the high density of warm particles. We believe this occurs when passing particles equilibrate with a population of particles which are neither absolutely trapped nor passing, that exist in a chaotic region around the loss cone which has previously been described. 

\section*{Acknowledgements}
 We are grateful to Princeton Fusion Systems for their support of C. Galea. This work was supported by the U.S. Department of Energy under contract number DE-AC02-09CH11466. The United States Government retains a non-exclusive, paid-up, irrevocable, world-wide license to publish or reproduce the published form of this manuscript, or allow others to do so, for United States Government purposes.
\section*{Data Availability Statement}
The figures in this paper are openly available.

\bibliographystyle{aiaa}
\bibliography{references}

\end{document}